\documentclass[twocolumn,showpacs,floatfix]{revtex4}%
\usepackage{graphicx}%
\usepackage{amsmath}%
\usepackage{amsfonts}%
\usepackage{amssymb}
\usepackage{bm}

\def\d{{\partial}}
\def\s{{\sigma}}
\def\e{{\epsilon}}
\def\k{{ {\bm k} }}
\def\p{{ {\bm p} }}

\def\w{{\omega}}
\def\a{{\alpha}}
\def\b{{\beta}}

\allowdisplaybreaks[4]

\begin{document}
\title{Effect of realistic finite-size impurities on $T_{\rm c}$
in Fe-based superconductors \\
based on the five-orbital tight-binding model
}
\author{Youichi \textsc{Yamakawa}$^{1}$,
Seiichiro \textsc{Onari}$^{2}$,
and Hiroshi \textsc{Kontani}$^{1}$
}

\date{\today }

\begin{abstract}
%We present a quantitative study of the reduction in $T_{\rm c}$
%($\Delta T_{\rm c}$)
%and residual resistivity ($\rho_0$) in Fe-based superconductors,
%by introducing realistic impurity models of Fe-site substitution.
We present a systematic study of the impurity effect on 
$T_{\rm c}$ in Fe-based superconductors,
assuming that the sign-reversal $s$-wave state due to 
inter-pocket repulsion ($s_\pm$-wave state) is realized.
For this purpose, we introduce several realistic impurity models
with non-local modifications of potentials
and hopping integrals around the impurity site.
When we use the impurity model parameters for $3d$- and $4d$-impurity atoms
derived from the recent first principle study by Nakamura {\it et al}.,
%in which the on-site impurity potential $I$ gives the largest scattering,
we find that the $s_\pm$-wave state is very fragile against impurities:
The superconductivity without impurities $T_{\rm c0}=30K$ 
is destroyed by introducing small residual resistivity 
$\rho_{0}^{\rm cr}=5z^{-1}\sim10z^{-1} \ [\mu\Omega {\rm cm}]$
($z^{-1}=m^*/m$ being the mass-enhancement factor),
consistently with the previous theoretical study for the 
on-site impurity model by Onari and Kontani.
This result is essentially unchanged for different 
non-local impurity models with realistic parameters.
We also discuss the effect of the impurity-induced 
non-local orbital order on the superconducting state.

%Even if the nearest-neighbor impurity potential $V_1$
%gives the dominant scattering, $\rho_{0}^{\rm cr}$ can increase to 
%at most $3\sim4$ times the value due to the on-site impurity $I$.
%$\sim 20z^{-1} \ {\rm K}/\Omega {\rm cm}$ for $T_{\rm c0}=30K$.
%Thus, the $s_\pm$-wave state is weak against
%realsitic finite radius impurities.

% against impurities 
%reported by Onari and Kontani is essentially correct even if the finite radius effect is taken into accout.

%We present a systematic study of impurity effect due to 
%various non-local impurity potentials, and 

%both possibilities in real Fe-based superconductors.
\end{abstract}
%Zhang,Sawadowski

\address{
$^1$ Department of Physics, Nagoya University,
Furo-cho, Nagoya 464-8602, Japan. 
\\
$^2$ Department of Applied Physics, Nagoya University,
Furo-cho, Nagoya 464-8602, Japan. 
}
 

\sloppy

\maketitle

%%%%%%%%%%%%%%%%%%
%Introduction
%%%%%%%%%%%%%%%%%%
\section{Introduction}

Since the discovery of Fe-based high-$T_{\rm c}$ superconductors
\cite{Hosono},
the symmetry and the gap structure of the superconducting (SC) state
have been studied very intensively.
It had been established experimentally that 
$s$-wave ($A_{1g}$ symmetry) SC state
is realized in usual Fe-based superconductors.
The gap structure in many optimally-doped high-$T_{\rm c}$ compounds
is nearly isotropic and fully-gapped
 \cite{Hashimoto,Tanatar,Hirsch-rev},
although some compounds show accidental nodal gap structure.
In the phase diagram, the SC phase is realized next to the 
orthorhombic structure transition at $T_{\rm S}$,
and the magnetic order is also realized at $T_{\rm N}\lesssim T_{\rm S}$.
Below $T_{\rm S}$, 
the orbital polarization $n_{xz}\ne n_{yz}$ is realized
\cite{ARPES-Shen},
and sizable softening of shear modulus $C_{66}$
\cite{Fernandes,Yoshizawa,Goto}
indicates the development of orbital fluctuations near the 
orthorhombic phase.
Strong spin fluctuations are also observed near the 
magnetic ordered phase.

These observed orbital and spin quantum criticalities
have been intensively studied theoretically,
since they would be closely connected to the pairing mechanism.
Within the random-phase-approximation (RPA),
strong spin fluctuations develop in
the multiorbital Hubbard models for Fe-based superconductors.
Therefore, spin-fluctuation-mediated $s$-wave state with sign reversal 
($s_\pm$-wave state) is obtained by the RPA
\cite{Kuroki,Mazin,Hirschfeld,Chubukov}.
However, the RPA fails to explain the 
non-magnetic structure transition at $T_{\rm S}$.
Also, orbital-fluctuation-mediated $s$-wave state 
without sign reversal ($s_{++}$-wave state) is realized by 
introducing the quadrupole interaction $g_{\rm quad}$ 
due to Fe-ion oscillations
\cite{Kontani-RPA,Saito-RPA,Saito-RPA2}.
Even for $g_{\rm quad}=0$, strong orbital fluctuations 
are obtained by improving the RPA by including the vertex correction (VC)
for the susceptibility that is dropped in the RPA 
\cite{Onari-SCVC,Ohno-SCVC}:
Since spin and orbital fluctuations mutually develop in
the self-consistent VC %(SC-VC) 
method, both $s_{\pm}$-wave and $s_{++}$-wave states
can be obtained by solving the multiorbital Hubbard model.

To distinguish between these two SC gap states,
various phase-sensitive experiments had been performed
 \cite{Hanaguri,christianson,keimer,tate,Sato-imp,Nakajima,Li2,Paglione}.
For example, inelastic neutron scattering experiments 
had been performed to find the magnetic resonance scattering
due to the sign reversal \cite{christianson,keimer,tate}.
However, observed ``resonance-like'' hump structure 
can be explained even if $s_{++}$-wave state is realized 
\cite{Onari-resonance}.
Impurity effect measurement is another significant
phase-sensitive experiment
since $T_{\rm c}$ would be strongly suppressed by 
inter-pocket impurity scattering of Cooper pairs
if $s_\pm$-wave state is realized
\cite{Sato-imp,Nakajima,Li2,Paglione}.
Many theoretical studies have been devoted so far
\cite{Chubukov,Senga1,Bang,Senga2,Onari-imp,FCZhang,Hirsch-imp}.
In Ref. \cite{Onari-imp},
the present authors studied the impurity effect due to 
local nonmagnetic impurities based on the realistic multiorbital model,
and showed that $T_{\rm c}$ in the $s_\pm$-wave state is 
strongly suppressed by inter-pocket impurity scattering of Cooper pairs.
However, effect of the possible non-locality of the impurity potential
had not been studied.

%The reduction in $T_{\rm c}$ per residual resistivity is
%$-\Delta T_{\rm c}/\rho_0 \sim 5z \ {\rm K}/\Omega {\rm cm}$,
%where $z^{-1}=m^*/m$ is the mass enhancement factor,
%independently of the impurity potential strength.

Therefore, in this paper, we present a quantitative study of the 
{\it non-local} impurity effect on the $s_\pm$-wave state
based on the realistic five-orbital model for Fe-based superconductors.
We introduce realistic models of Fe-site substitutional impurity atoms,
which contains the non-local modifications of potentials ($I,V_1,V_2$)
and hopping integrals ($\Delta t$) around the impurity site,
referring to a recent first principle study by Nakamura {\it et al}.
 \cite{Nakamura}.
%When the on-site impurity potential $I$
%gives the dominant scattering, 
%which corresponds to the Fe-site substitution by other atoms,
In various non-local impurity models with realistic parameters,
the $s_\pm$-wave state is very fragile against impurities:
Original transition temperature at $T_{\rm c0}=30K$ is destroyed
when the residual resistivity is just
$\rho_{0}^{\rm cr}=5z^{-1}\sim 10z^{-1} \ [\mu\Omega {\rm cm}]$,
where $z^{-1}=m^*/m$ is the mass-enhancement factor.
This result is consistent with the previous theoretical study 
for the on-site impurity model in Ref. \cite{Onari-imp}.
Thus, experimentally observed robustness of $T_{\rm c}$
against impurities in various (optimally-doped) Fe-based superconductors
\cite{Sato-imp,Nakajima,Li2,Paglione}
would indicate the realization of the $s_{++}$-wave state.

The effect of Fe-site substitution on $T_{\rm c}$ 
had been carefully studied in La(Fe,$X$)As(O,F) for $X$=Co,Ni
\cite{Kawamata}
and in Ba(Fe,$X$)$_2$As$_2$ as well as Ba(Fe$_{1-x-y}$Co$_x$Cu$_y$)$_2$As$_2$
\cite{Canfield1,Canfield2,Ideta}.
In these materials, $T_{\rm c}$ is well scaled by the 
amount of carrier doping caused by $X$-atom substitutions,
not by the impurity concentration
irrespective of the large impurity potential \cite{Ideta}.
This fact would mean the robustness of the SC state 
against strong impurity scattering in these materials 
as stressed in Ref. \cite{Kawamata}.

In Fe-based superconductors,
%intra-band and inter-band 
impurity potential matrix ${\hat I}_{\rm b}(\k,\k')$
in the {\it band-diagonal basis}
is $\k$-dependent, reflecting the multiorbital band structure \cite{Onari-imp}.
However, their $\k$-dependences had been frequently 
neglected in previous studies for simplicity.
In this ``constant ${\hat I}_{\rm b}$ model'',
both the intra-band and inter-band scatterings, $I_{\rm b}$ and $I'_{\rm b}$,
are constant parameters.
In this model, the $s_\pm$-wave state can be stable against impurities when 
$|I'_{\rm b}/I_{\rm b}|\ll1$, although it seems unrealistic
since both hole-Fermi surfaces (h-FSs) and 
electron-Fermi surfaces (e-FSs) are composed of
the common $d$-orbitals \cite{Onari-imp}.
In addition, obtained results are strongly changed in the unitary 
and intermediate regimes
once the $\k$-dependence of ${\hat I}_{\rm b}$ is taken into account;
see Sec. \ref{sec:Senga-dis}.
%This artifact is resolved by taking the
%momentum dependence of ${\hat I}_{\rm b}$
%into account correctly, based on the multiorbital model 
%\cite{Onari-imp}. 
%Therefore, 
%realistic impurity models with finite radius have to be analyzed
%based on the realistic multiorbital model.

%%%%%%%%%%%%%%%%%%
\section{Model Hamiltonian}
\label{sec:model}
In this paper,
we study the impurity effect based on the realistic two-dimensional
five-orbital tight-binding model 
\cite{Kuroki}:
\begin{eqnarray}
H_{0} = \sum_{\k,\s,l,l'} h_\k^{l,l'}c_{\k l \s}^\dagger c_{\k l' \s} ,
\end{eqnarray}
where ${\hat h}_\k$ is the $5\times5$ matrix
given by the Fourier transformation of the hopping integral
$t_{{\bm r}l,{\bm r}'l'}^0$ introduced in Ref. \cite{Kuroki}.
Here, $l,l'$ represents the orbital indices, and $\s$ is the spin index.
The matrix elements of ${\hat h}_\k$ 
is given by the Fourier transformation of the 
hopping integral $t_{{\bm r}l,{\bm r'}l'}^0$.
When the electron filling per Fe-site is $n=6.0$,
There are two hole-pockets around the $\Gamma$ point,
one hole-pocket around $(\pi,\pi)$ point,
and two electron-pockets around $(\pi,0)$ and $(0,\pi)$ points.

\begin{figure}[!htb]
\includegraphics[width=0.50\linewidth]{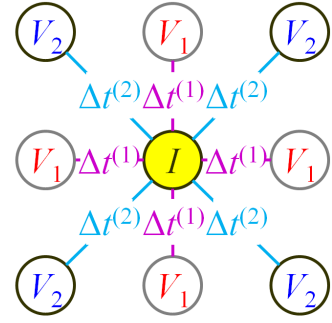}
\caption{
(Color online)
The present non-local impurity potential.
}
\label{fig:Imp-realspace}
\end{figure}

In addition, we introduce the following nonmagnetic and 
non-local impurity potential at site ${\bm 0}=(0,0)$:
\begin{eqnarray}
H_{\rm imp}&=& I\sum_{l,\s} c_{{\bm 0}l\s}^\dagger c_{{\bm 0}l\s}
\nonumber \\
& &+ V_1 \sum_{\bm r}^{\rm NN}\sum_{l,\s} c_{{\bm r}l\s}^\dagger  c_{{\bm r}l\s}
   + V_2 \sum_{{\bm r}'}^{\rm NNN}\sum_{l,\s} c_{{\bm r}'l\s}^\dagger  c_{{\bm r}'l\s}
\nonumber \\
& &+ \sum_{\bm r}^{\rm NN}\sum_{l,l',\s} 
\Delta t_{{\bm 0}l,{\bm r}l'}^{(1)} (c_{{\bm 0}l\s}^\dagger  c_{{\bm r}l'\s} + {\rm h.c.})
\nonumber \\
& &+ \sum_{\bm r'}^{\rm NNN}\sum_{l,l',\s} 
\Delta t_{{\bm 0}l,{\bm r'}l'}^{(2)} (c_{{\bm 0}l\s}^\dagger  c_{{\bm r'}l'\s} + {\rm h.c.})
\nonumber \\
&\equiv&
\sum_{{\bm r},{\bm r'}}\sum_{l,l',\s} W_{{\bm r}l,{\bm r'}l'}
c_{{\bm r}l\s}^\dagger  c_{{\bm r'}l'\s} ,
 \label{eqn:Himp}
\end{eqnarray}
where $I$, $V_1$ and $V_2$ are the on-site, nearest-neighbor (NN),
and next-nearest-neighbor (NNN) impurity potential,
and $\Delta t_{{\bm 0}l,{\bm r}l'}^{(i)}$ is the 
modulation of the NN or NNN hopping integrals between 
site ${\bm 0}$ and site ${\bm r}$.
The present impurity potential model
is depicted in Fig. \ref{fig:Imp-realspace}.

%%%%%%%%%%%%%%%%%%%%%%%%%%%%%%%%%
%%%%%%%%%%%%%%%%%%%%%%%%%%%%%%%%%

%%%%%%%%%%%%%%%%%%%%%%%%%%%%%%%%
\section{gap equation, calculation of residual resistivity}
\label{sec:T-matrix}
%Now, we represents eq. (\ref{eqn:Himp}) as
%$H_{\rm imp}=\sum_{{\bm r},{\bm r'}}\sum_{l,l',\s} W_{{\bm r},{\bm r'}}^{l,l'}
%c_{{\bm r},l,\s}^\dagger  c_{{\bm r'},l',\s}$.
In the present model, the $T$-matrix due to infinite number of 
impurity scattering processes is given as
\begin{eqnarray}
{\hat T}(i\e_n)= {\hat W}(1-{\hat g}(i\e_n)\cdot{\hat W})^{-1} ,
\label{eqn:T-def}
\end{eqnarray}
where ${\hat g}_{{\bm r}l,{\bm r'}l'}(i\e_n)$ is the 
free Green function in real space, given by the Fourier transformation
of ${\hat G}_\k(i\e_n)=(i\e_n+\mu-{\hat h}_\k)^{-1}$, and 
$\e_n=(2n+1)\pi T$ is the fermion Matsubara frequency.

When the impurity concentration is dilute ($n_{\rm imp}\ll1$),
the normal self-energy above $T_{\rm c}$ is well approximated by 
the $T$-matrix approximation.
It is given by
\begin{eqnarray}
\delta{\hat \Sigma}^n_\k(i\e_n)= \left.n_{\rm imp}{\hat T}_{\k,\k'}(i\e_n)\right|_{\k=\k'} ,
\label{eqn:NSigma}
\end{eqnarray}
where ${\hat T}_{\k,\k'}(i\e_n)$ is given by the Fourier transformation of 
eq. (\ref{eqn:T-def}).
The retarded (advanced) self-energy is given by the 
analytic continuation $i\e_n \rightarrow \e+i\delta\cdot {\rm sgn}(n)$:
$\delta{\hat \Sigma}_\k^{n\rm R(A)}(\e)=
\delta{\hat \Sigma}(\k,\e+(-)i\delta)$.
Then, the Green function in the band-diagonal basis is
$G_{\a\k}^{\rm R}(\e)= 1/(\e+\mu-E_{\a\k}-i\gamma_{\a\k}(\e))$,
where $\a$ is the band basis, $E_{\a\k}$ is the 
dispersion of the $\a$th band, and
$\gamma_{\a\k}(\e)=-{\rm Im}\delta{\hat \Sigma}_\k^{n \rm R}(\e)$
is the quasiparticle damping rate due to impurities.
$\gamma_{\a\k}(0)$ can be rewritten as
\begin{eqnarray}
\gamma_{\a\k}(0)&=& 
-n_{\rm imp}\sum_{\k',\b} |T_{\a\k,\b\k'}^R(0)|^2 {\rm Im}G_{\b\k'}^R(0)
 \nonumber \\
&=& \frac{n_{\rm imp}}{4\pi}
\sum_{\b}\int_{\rm FS\b}\frac{d\k'}{v_{\b\k'}} |T_{\a\k,\b\k'}^R(0)|^2 
\label{eqn:gamma} ,
\end{eqnarray}
where $\int_{\rm FS\a} dk$ is the integration on the FS$\a$.

When the impurity concentration is low enough,
the inter-band contribution to the conductivity is negligible.
If we drop the current vertex correction (CVC),
the conductivity is given by
\begin{eqnarray}
\s_\nu^{\rm no \ CVC}= \frac{e^2}{c}\sum_{\k,\a}^{\rm band}\int \frac{d\e}{\pi}
\left(-\frac{\d f}{\d\e}\right)_{E_{\a\k}}
|G_{\a\k}^{\rm R}(\e)|^2 v_{\a\k,\nu}^2 ,
\end{eqnarray}
where $\nu=x$ or $y$,
$f(\e)=(e^{\e/T}+1)^{-1}$, $v_{\a\k,\nu}= d E_{\a\k}/d k_\nu$
is the quasiparticle velocity, and $c$ is the inter-layer distance.
However, to obtain the exact conductivity for $n_{\rm imp}\ll1$,
the CVC should be taken into account.
The exact expression for the conductivity is given as
\begin{eqnarray}
\s_\nu= \frac{e^2}{c}\sum_{\k,\a}^{\rm band}\int \frac{d\e}{\pi}
\left(-\frac{\d f}{\d\e}\right)_{E_{\a\k}}
|G_{\a\k}^{\rm R}(\e)|^2 v_{\a\k,\nu} J_{\a\k,\nu}(\e) .
\end{eqnarray}
Hereafter, we put $c=0.6$nm that corresponds to Ba122 compounds.
$J_{\a\k,\nu}$ is the total velocity with CVC, which is 
given by solving the following Bethe-Salpeter equation:
\begin{eqnarray}
J_{\a\k,\nu}(\e)&=& v_{\a\k,\nu} 
\nonumber \\
& & \!\!\!\!\!\!\!\!\!\!\!
+n_{\rm imp}\sum_{\k',\b}|T_{\a\k,\b\k'}^R(\e)|^2 
%\nonumber \\
%& &\ \ \ \ \ \ \ \ \ \ \times 
|G_{\b\k'}^{\rm R}(\e)|^2 J_{\b\k',\nu}(\e) ,
\label{eqn:BS}
\end{eqnarray}
where $T_{\a\k,\b\k'}^R(\e)$ is the $T$-matrix in the band-diagonal basis.
Here, we neglect the inter-band contribution
since it is negligible when $n_{\rm imp}\ll 1$.

Next, we study the impurity effect on $T_{\rm c}$.
The gap equation at $T=T_{\rm c}$ in the band basis is given by
\begin{eqnarray}
Z_{\a\k}(\e_n)\Delta_{\a\k}(\e_n)&=& - \frac{T_{\rm c}}{4\pi} 
\sum_{\b,m} \int_{\rm FS\b} \frac{d\k'}{v_{\b\k'}}V_{\a\k,\b\k'}(\e_n,\e_m) 
\nonumber \\
& &\times \frac{\Delta_{\k'}^\b(\e_m)}{|\e_m|}
+\delta\Sigma_{\a\k}^a(i\e_n) ,
\label{eqn:gap-eq}
\end{eqnarray}
where 
$\Delta_{\a\k}$ is the SC gap function in the band-diagonal basis, 
and $V$ is the pairing interaction.
$Z_{\a\k}(\e_n)= 1+\gamma_{\a\k}(\e_n)/|\e_n|$, and
$\delta\Sigma^a_{\a}$ is the linearized anomalous self-energy is given as
\begin{eqnarray}
\delta {\hat \Sigma}^a_{\a\k}(i\e_n)&=& n_{\rm imp}
\sum_{\k',\b} |T_{\a\k,\b\k'}(i\e_n)|^2 |G_{\b\k'}(i\e_n)|^2
\Delta_{\b\k'}(i\e_n )
 \nonumber \\
&=& \frac{n_{\rm imp}}{4\pi}
\sum_{\b}\int_{\rm FS\b}\frac{d\k'}{v_{\b\k'}} |T_{\a\k,\b\k'}(i\e_n)|^2 
\frac{\Delta_{\b\k'}(i\e_n )}{|\e_n|} ,
 \nonumber \\
\label{eqn:ASigma}
\end{eqnarray}
By solving the gap equation (\ref{eqn:gap-eq})
together with eqs. (\ref{eqn:NSigma}) and (\ref{eqn:ASigma}),
the impurity effect on $T_{\rm c}$ for $n_{\rm imp}\ll1$
is exactly calculated within the BCS theory.

The reduction in $T_{\rm c}$ is caused by the function
$\gamma_{\a\k}\cdot \Delta_{\a\k}/|\e_n|-\delta {\hat \Sigma}^a_{\a\k}$,
which vanishes if $\Delta_{\a\k}$ is independent of band and momentum
(=isotropic $s_{++}$-wave state)
according to eqs. (\ref{eqn:gamma}) and (\ref{eqn:ASigma}).
Then, the independence of $T_{\rm c}$ against impurities
is derived from this relation, known as the Anderson theorem.
However, Anderson theorem is totally violated in the $s_\pm$-wave state.

In the present study,
we consider the case of spin fluctuation mediated
intra-orbital repulsive interaction between e-FS and h-FS.
Then, the spin susceptibility is approximately 
diagonal with respect to the $d$-orbital,
since it is strongly enhanced by the intra-orbital repulsion $U$.
That is, spin-spin correlation develops only in the same $d$ orbital.
By taking this fact into account, we introduce the following 
repulsive pairing interaction:
\begin{eqnarray}
V_{\a\k,\b\k'}(\e_n,\e_{n'})&=& g\sum_l^{d-{\rm orbital}} 
|U_{l,\a}(\k)|^2|U_{l,\b}(\k')|^2
\nonumber \\
& &\times D(\e_n)D(\e_{n'}) ,
\label{eqn:V}
\end{eqnarray}
for $\a\in$e-FS and $\b\in$h-FS and vise versa, 
and $V_{\a\k,\b\k'}=0$
for $\a,\b \in$e-FS or $\a,\b\in$h-FS.
$U_{l,\a}(\k)=\langle l,\k|\a,\k \rangle$ is the unitary matrix 
connecting between band-basis and orbital-basis, and
$D(x)=\w_c^2/(x^2+\w_c^2)$, where
$\w_c$ is the cut-off energy of the pairing interaction.
In the present study, we put $g=3.22$eV and $\w_{\rm c}=0.03$eV.
In this case, a fully-gapped $s_\pm$-wave state 
with $T_{\rm c0}=30$K is realized when $n_{\rm imp}=0$.
Hereafter, we set the unit of energy is eV, 
and the unit of temperature is K.

%%%%%%%%%%%%%%%%%%%%%%%%%%%%%%%%
\section{numerical results}
\label{sec:numerical}

%%%%%%%%%%%%%%%%%%%%%%%%%%%%%%%
\subsection{Nakamura's impurity model}
\label{sec:numerical-Nakamura}

First, we study the non-local impurity potential
$W_{{\bm r}l,{\bm r'}l'}^{X}$ for $X$-atom
($X$= Mn, Co, Ni, Zn, Ru) given by the 
first principle study by Nakamura {\it et al.} \cite{Nakamura}.
The derived values of the on-site impurity potential $I$ are
$+0.28$, $-0.35$, $-0.87$, $-8.05$ and $-0.02$ (the unit is eV)
for Mn, Co, Ni, Zn and Ru, respectively,
as shown in Table \ref{tab:table1}.
Thus, Zn atom works as the unitary impurity scattering center,
whereas other 3$d$ atoms (Mn, Co and Ni) induce
intermediate impurity scattering.
The off-site impurity potentials ($V_1$ and $V_2$) are very small.
In addition, strong modulations of NN and NNN hopping integrals
($\Delta t^{(1)}$ and $\Delta t^{(2)}$)
are induced around the impurity atom.
Especially, $\Delta t^{(i)}/t^0$ for $4d$ atom Ru takes
large positive value, reflecting the larger radius of 4$d$-orbitals.
Thanks to the Nakamura's model, we can present a quantitative analysis 
of the impurity effect on the $s_\pm$-wave state,
by taking the non-local nature of the impurity potential appropriately.
%beyond a simple on-site impurity model.
% \cite{Onari-imp}
 
%%%%%%%%%%%%%%%%%%%%%%%%%%%%%%%%%%%%%%%%%%%%%%
\begin{table}[h]
\begin{center}
\begin{tabular}{|c|c|c|c|c|c|}   
\hline
 & Mn & Co & Ni & Zn & Ru \\
\hline
$I_{\rm Ave}$ & $+0.28$ & $-0.35$ & $-0.87$ & $-8.05$ & $-0.02$ \\
\hline
$\Delta N(I_{\rm Ave})$ & $-1.14$ & $+1.09$ & $+2.05$ & $+3.91$ & $-0.02$ \\
\hline
\end{tabular}
\caption{\label{tab:table1}
$I_{\rm Ave}$ is the averaged on-site impurity potential in
Nakamura's model \cite{Nakamura}.
The unit of energy is eV.
$\Delta N(I_{\rm Ave})$ is the change in the local electron density 
at the impurity site due to $I_{\rm Ave}$ in the present model
without interaction.
In Nakamura's model, the on-site potential is orbital-dependent,
and $\Delta t^{(1)}$ and $\Delta t^{(2)}$ are included.
%$I$ is the  on-site impurity potential 
%that produce the change in the local electron density
%by integer ($\Delta N(I)$). 
}
\end{center}
\end{table}
%%%%%%%%%%%%%%%%%%%%%%%%%%%%%%%%%%%%%%%%%%%%%%

\begin{figure}[!htb]
\includegraphics[width=0.99\linewidth]{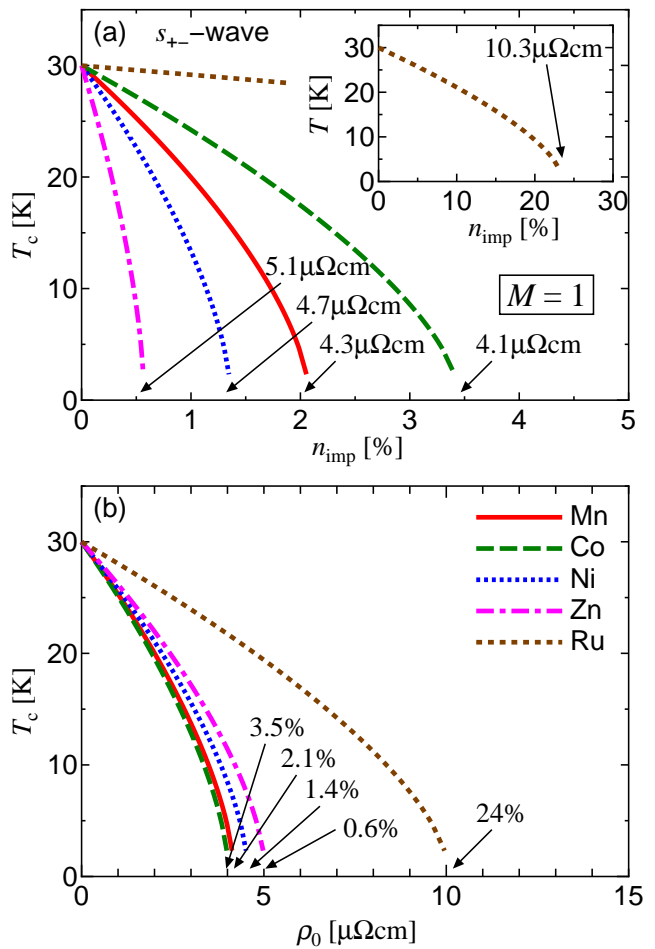}
\caption{
(Color online)
Obtained $T_{\rm c}$ as function of (a) $n_{\rm imp}$ and (b) $\rho_0$
using the impurity potential $W_{{\bm r}l,{\bm r'}l'}^{X}$ 
for $X$-atom ($X$= Mn, Co, Ni, Zn, Ru).
}
\label{fig:Nakamura-Tc-nimp}
\end{figure}

Figure \ref{fig:Nakamura-Tc-nimp} 
show the obtained $T_{\rm c}$ as function of the
(a) impurity concentration $n_{\rm imp}$ and 
(b) residual resistivity $\rho_0=1/\s_x$
for various impurity atoms in the case of $z^{-1}=1$.
Although the Fe-site substitution induces the 
``impurity potential'' and ``carrier doping'', 
we neglect the latter effect by fixing the electron filling $n=6.0$, 
in order to concentrate on the former effect.
When the mass-enhancement factors $z^{-1}$ is finite,
the reduction of $T_{\rm c}$ per impurity concentration, 
$-(T_{\rm c}-T_{\rm c0})/n_{\rm imp}$, is renormalized by $z$,
while $\rho_0/n_{\rm imp}$ is independent of $z$ \cite{ROP}.
Therefore, both $n_{\rm imp}^{\rm cr}$ and 
$\rho_0^{\rm cr} \equiv \rho_0(n_{\rm imp}^{\rm cr})$ are 
multiplied by $z^{-1}$.
According to Fig.  \ref{fig:Nakamura-Tc-nimp} (a),
the critical impurity concentration for the disappearance of $T_{\rm c}$,
$n_{\rm imp}^{\rm cr}$, strongly depends on the impurity atoms:
$n_{\rm imp}^{\rm cr}=0.6z^{-1}\ [\%] \sim 3.5z^{-1}\ [\%]$ 
for 3$d$-impurities (Mn, Co, Ni, Zn),
while $n_{\rm imp}^{\rm cr}=24z^{-1}\ [\%]$ for Ru-impurities.
In contrast, the values of $\rho_0^{\rm cr}$ shown in 
Fig. \ref{fig:Nakamura-Tc-nimp} (b) are almost independent of impurities
for $3d$-impurity atoms ($\sim5 z^{-1} \ [\mu\Omega{\rm cm}]$),
while $\rho_0^{\rm cr} \sim10 z^{-1} \ [\mu\Omega{\rm cm}]$ for
Ru-impurity.

%%%%%%%%%%%%%%%%%%%%%%%%%%%%%%%%%
%%%%%%%%%%%%%%%%%%%%%%%%%%%%%%%%%

\begin{figure}[!htb]
\includegraphics[width=0.99\linewidth]{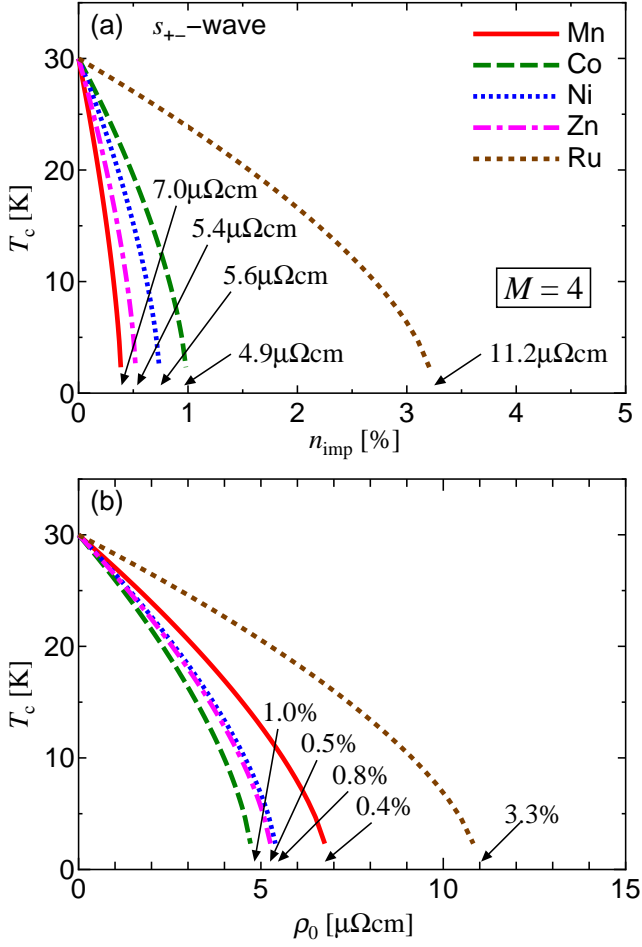}
\caption{
(Color online)
Obtained $T_{\rm c}$ as function of (a) $n_{\rm imp}$ and (b) $\rho_0$
using the impurity potential $4\times W_{{\bm r}l,{\bm r'}l'}^{X}$.
}
\label{fig:Nakamura-Tc-nimp2}
\end{figure}

The residual resistivity given by Nakamura parameter $W_{{\bm r}l,{\bm r'}l'}^{X}$ 
is very small ($\rho_0/n_{\rm imp}<5\ [\mu\Omega{\rm cm}/\%]$)
expect for $X={\rm Zn}$.
One of the reasons would be that the
impurity potential given in Ref. \cite{Nakamura} may be normalized,
although the bare impurity potential is required for the present study.
(The normalization would caused by the modification of
the wavefunction around the impurity site in solving the 
Kohn-Sham equation, while this process is also included in the $T$-matrix.)
By taking this effect into account, we study the impurity potential
$M\times W_{{\bm r}l,{\bm r'}l'}^X$ with $M>1$.
To see the effect of $M$, we analyze the case of $M=4$ 
in Fig. \ref{fig:Nakamura-Tc-nimp2} for instance.
We show (a) $n_{\rm imp}$-dependence and (b) $\rho_0$-dependence of $T_{\rm c}$.
Note that similar results are obtained for $M\ge 4$.
In this case, $T_{\rm c0}=30$K is suppressed by a few percent 
impurity concentration.
In contrast, $\rho_0^{\rm cr}$ is approximately independent of $M$
for all impurity atoms.
We stress that $\rho_0/n_{\rm imp}$ is enlarged to 
$\sim50 \ [\mu\Omega{\rm cm}/\%]$
by taking the impurity-induced non-local orbital order into account,
because of the enlarged impurity scattering cross section \cite{Inoue}.
In this case, $s_\pm$-wave state is strongly suppressed, 
as we will show in Sec. \ref{sec:Inoue}.

%Figure \ref{fig:Nakamura-Tc-rho} shows the 
%obtained $T_{\rm c}$'s as function of $\rho_0$ 
%for Nakamura's impurity model with $M=1$.
%The obtained results are very similar for Mn, Co, Ni and Zn,
%except for Ru impurity atoms.
%The obtained $T_{\rm c}$'s for $M>1$
%are very similar to Fig. \ref{fig:Nakamura-Tc-rho}.
%Therefore, the results in Fig. \ref{fig:Nakamura-Tc-rho},
%which means the weakness of the $s_\pm$-wave state
%against impurities, 
%are universal irrespective of the impurity potential strength.

%%%%%%%%%%%%%%%%%%%%%%%%%%%%%%%%%
%%%%%%%%%%%%%%%%%%%%%%%%%%%%%%%%%

\begin{figure}[!htb]
\includegraphics[width=0.9\linewidth]{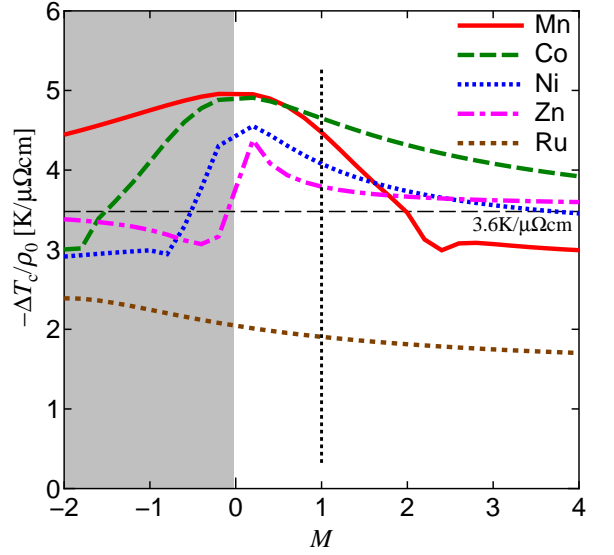}
\caption{
(Color online)
$R=-(T_{\rm c}-T_{\rm c0})/\rho_0$ at $n_{\rm imp}=0.1\ [\%]$ 
for the impurity potential $M\times W^X$ for various $X$ atoms.
The horizontal broken like at $R_{I=\infty}=3.6 \ [{\rm K}/\mu\Omega{\rm cm}]$
represents the value for the infinite on-site impurity potential ($I=\infty$)
studied in Ref. \cite{Onari-imp} for $z^{-1}=1$.
}
\label{fig:Nakamura-R-M}
\end{figure}

For the convenience of analysis, we hereafter study the ratio
$R\equiv -(T_{\rm c}-T_{\rm c0})/\rho_0 \equiv -\Delta T_{\rm c}/\rho_0$ 
at $n_{\rm imp}\ll n_{\rm imp}^{\rm cr}$:
$R$ is rather independent of the impurity potential strength, and also
$R$ is essentially independent of $T_{\rm c0}$
according to the Abrikosov-Gorkov theory.
In Fig. \ref{fig:Nakamura-R-M}, we show the obtained $R$
as function of $M$ for various impurity atoms.
As recognized in Ref. \ref{fig:Nakamura-Tc-nimp},
the relation $T_{\rm c0}/\rho_0^{\rm cr}\sim 1.5R$ is satisfied for all atoms.
By taking the mass-renormalization factor into account,
we obtain that $R=3z \sim 5z \ [{\rm K}/\mu\Omega{\rm cm}]$ 
for all $3d$-impurity atoms studied in Ref. \cite{Nakamura},
except for the Ru-impurity.
(The horizontal broken line $R_{I=\infty}=3.6 \ [{\rm K}/\mu\Omega{\rm cm}]$
is the value for the infinite on-site impurity potential ($I=\infty$)
studied in Ref. \cite{Onari-imp}.)

In contrast, experimentally observed $R$ in optimally-doped 
1111 compounds ($z^{-1}\sim2$) and 122 compounds ($z^{-1}\sim3$)
is $R_{\rm exp} \sim 0.1 \ [{\rm K}/\mu\Omega{\rm cm}]$ 
 \cite{Sato-imp,Li2,Paglione,Nakajima}.
Therefore, the $s_\pm$-wave state would be too fragile against
nonmagnetic impurities to explain experimental 
robustness of $T_{\rm c}$ against impurities.
Similar result was reported by 
Ikeda {\it et al} \cite{Ikeda}
by using the Nakamura's impurity model.

%%%%%%%%%%%%%%%%%%%%%%%%%%%%%%%%%
%%%%%%%%%%%%%%%%%%%%%%%%%%%%%%%%%

%%%%%%%%%%%%%%%%%%%%%%%%%%%%%%%
\subsection{ideal non-local impurity models}
\label{sec:numerical-V1}

\begin{figure}[!htb]
\includegraphics[width=0.6\linewidth]{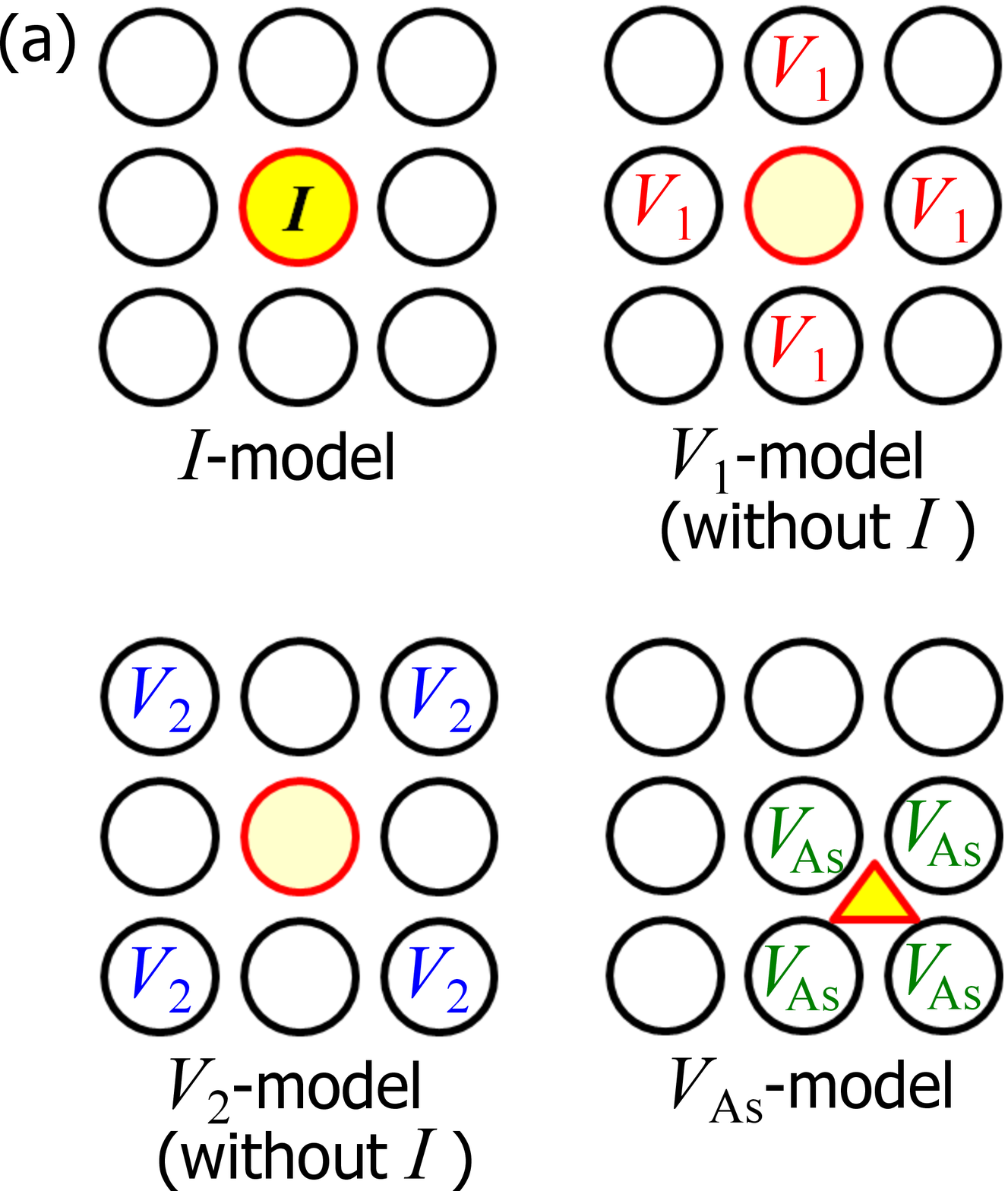}
\includegraphics[width=0.8\linewidth]{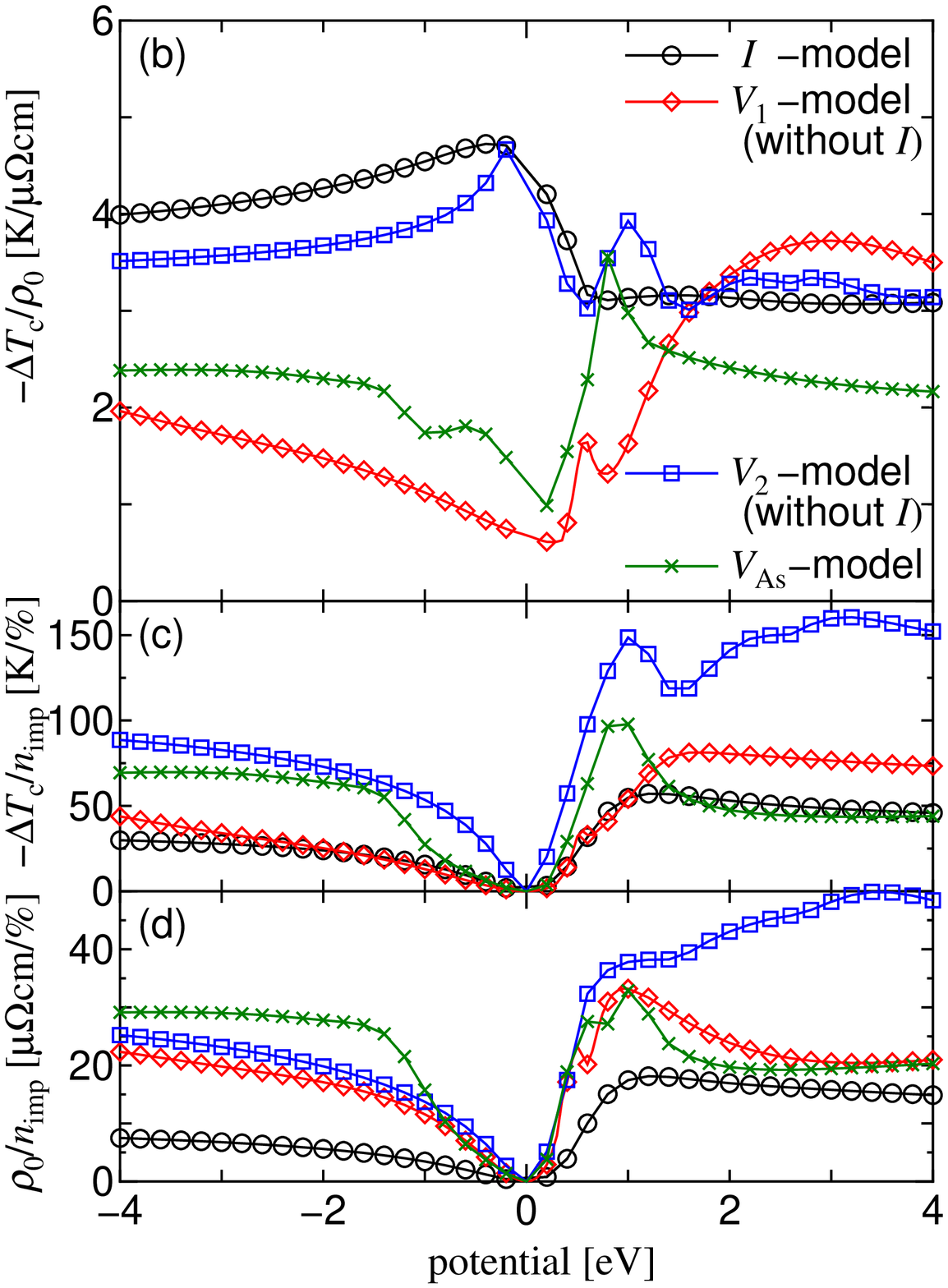}
\caption{
(Color online)
(a) Impurity potentials
for the $I$-model, $V_1$-model (without $I$), 
$V_2$-model (without $I$), and $V_{\rm As}$-model.
(b) Obtained $R=-(T_{\rm c}-T_{\rm c0})/\rho_0$ 
as function of the impurity potential.
Small $R$ for $V_1$-model is enlarged by introducing
the on-site potential $I$ if $|I|\gg|V_1|$; see Fig. \ref{fig:R-V1}.
(c) $-\Delta T_{\rm c}/n_{\rm imp}$  at $n_{\rm imp}=0.1\ [\%]$
and (d) $\rho_0/n_{\rm imp}$ as function of the impurity potential.
}
\label{fig:R-pot}
\end{figure}

According to the first principle studies in Refs. 
\cite{Nakamura,Zawa,Kemper},
the impurity-induced change in the electron density 
is strongly localized at the impurity center.
This fact indicates the smallness of the non-local impurity potentials $V_i$
in eq. (\ref{eqn:Himp}).
%large $V_i$ might be induced by inhomogeneities
%introduced by (for example) irradiations.
%Therefore, in the following, we will study the 
%effect of non-local potential $V_i$ on the $s_\pm$-wave state.
However, to obtain useful knowledge on the impurity effect,
we introduce four ideal impurity potential models
shown in Fig. \ref{fig:R-pot} (a):
(i) $I$-model (only on-site potential),
(ii) $V_1$-model (only NN potential, without $I$),
(iii) $V_2$-model (only NNN potential, without $I$), and 
(iv) $V_{\rm As}$-model (plaquette impurity potential 
due to As-site substitution).
Note that (ii) and (iii) are very unrealistic potentials.
Here, we put $\Delta t^{(i)}=0$ for simplicity.

Figure \ref{fig:R-pot} (b)
shows the ratio $R=-(T_{\rm c}-T_{\rm c0})/\rho_0$ 
at $n_{\rm imp}=0.1\ [\%]$ for models (i)-(iv) as function of 
the impurity potential, in the case of $z^{-1}=1$.
The obtained $R$ for the $V_2$-model is as large as that for $I$-model,
while those for other two models are smaller when $V_1,V_{\rm As}<0$.
Especially,  $R<1.0 \ [{\rm K}/\mu\Omega{\rm cm}]$
is realized  for the $V_1$-model for $-1<V_1<0.5$.
We also show $-\Delta T_{\rm c}/n_{\rm imp}$ and $\rho_0/n_{\rm imp}$ 
with CVC in Figs.  \ref{fig:R-pot} (c) and (d), respectively.
Compared to the $I$-model,
$-\Delta T_{\rm c}/n_{\rm imp}$ in the $V_1$ model is comparable,
while $\rho_0$ in the $V_1$ model is much larger.
For this reason, the relation $R_{I-{\rm model}}\gg R_{V_1-{\rm model}}$ 
can be achieved.

%We found that $R\sim R_0$ for $|V_1|\lesssim 0.2$,
%in the case of $|I|>4|V_1|$.
%On the other hand, we obtain
%$R\lesssim 0.3R_0$ for $V_1<-0.4$.
%Especially, $R\lesssim 0.1R_0$ for any $V_1$ in case of $I=0$.

%%%%%%%%%%%%%%%%%%%%%%%%%%%%%%%%%
%%%%%%%%%%%%%%%%%%%%%%%%%%%%%%%%%

%In Fig. \ref{fig:esult2-ratio}, we also show the ratio $R$
%for (c) $V_2\ne0$ ($I=V_1=0$) and (d) praquette impurity 
%potential $V_P$ at the rearest four sites.
%The latter impurity potential would be caused by 
%the As-ion substutution.
%In both cases, the relation $R\lesssim 0.5R_0$ is satisfied.

\begin{figure}[!htb]
\includegraphics[width=0.8\linewidth]{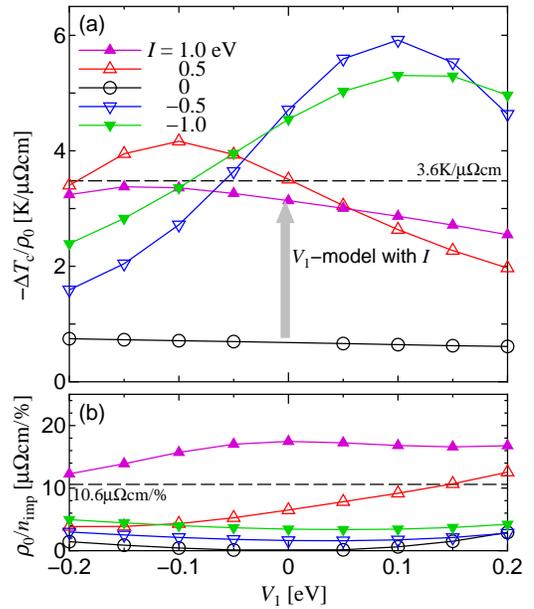}
\caption{
(Color online)
(a) $R$ for the $V_1$-model with on-site potential $I$.
For finite $|I| \ (\ge0.5)$,
$R\sim R_{I=\infty}=3.6 \ [{\rm K}/\mu\Omega{\rm cm}]$ 
is realized for $|V_1|\sim0.1$.
(b) The residual resistivity $\rho_0/n_{\rm imp}$.
The value $\rho_0^{I=\infty}/n_{\rm imp}=10.6\ [\mu\Omega{\rm cm}/\%]$
is shown by broken line.
%When $|I|\ge0.5$, $R\sim R_{I=\infty}$ is realized for $|V_1|\sim0.1$.
}
\label{fig:R-V1}
\end{figure}

However, the above $V_1$-model without on-site potential $I$ 
is very unrealistic.
Thus, we also introduce $I$ to the $V_1$-model:
Figure \ref{fig:R-V1} shows the obtained (a) $R$ and (b) $\rho_0/n_{\rm imp}$
as function of $V_1$ with finite $I$.
In the case of $|I|\ge0.5$,
$-\Delta T_{\rm c}/\rho_0^{\rm cr}$ quickly approaches to the 
value for the $I$-model for $|V_1|\lesssim 0.2$.
%The small residual resistivity in the $V_1$-model (without $I$)
%drastically increases with  $|I|$, while 
%$-\Delta T_{\rm c}/n_{\rm imp}$ is rather independent of $I$.
%For this reason, in the case of $|V_1|\lesssim 0.2$,
%$R$ is strongly enhanced by $|I|\gtrsim 0.5$.
%In the case of $|V_1|\gtrsim 0.4$, in contrast,
%$R$ could be suppressed to $\sim R_{I=\infty}/3$ even when $|I|\sim1$.
In real compound, the relation $|V_1|<0.1$ is expected,
as we will discuss in Sec. \ref{sec:V1-dis}.
%We will estimate of the value of $V_1$ in real materials in discussion.

%%%%%%%%%%%%%%%%%%%%%%%%%%%%%%%%%
%%%%%%%%%%%%%%%%%%%%%%%%%%%%%%%%%

We also study the effect of hopping integral inhomogeneity
around the impurity site.
For a systematic study, we define $\Delta t^{(i)}$ in 
eq. (\ref{eqn:Himp}) as
$\Delta t_{{\bm 0}l,{\bm r}l'}^{(1)}\equiv x_1\cdot t_{{\bm 0}l,{\bm r}l'}^0$
for the NN sites ($|{\bm r}|=1$), and 
$\Delta t_{{\bm 0}l,{\bm r}l'}^{(2)}\equiv x_2\cdot t_{{\bm 0}l,{\bm r'}l'}^0$
for the NNN sites ($|{\bm r'}|=\sqrt{2}$).
In the case of Ru-impurity,
$x_1$ and $x_2$ are positive according to Ref. \cite{Nakamura},
because of the larger radius of $4d$-orbitals.
However, we also study the case $x_i<0$ since 
this situation might be realized by irradiations,
by shifting the Fe-ion position outside of the FeAs plane.

\begin{figure}[!htb]
\includegraphics[width=0.85\linewidth]{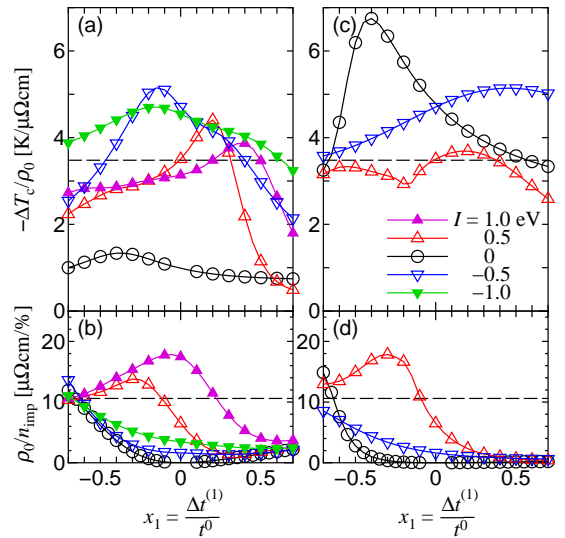}
\caption{
(Color online)
Obtained $R$ and $\rho_0/n_{\rm imp}$  as function of $x_1=\Delta t^{(1)}/t^0$,
in the case of (a)(b) $x_2=0$ and (c)(d) $x_2=x_1$.
$R\sim R_{I=\infty}=3.6 \ [{\rm K}/\mu\Omega{\rm cm}]$ 
is realized for $|I|\ge0.5$.
%The value $\rho_0^{I=\infty}/n_{\rm imp}=10.6\ [\mu\Omega{\rm cm}/\%]$
%is shown by brokenline.
}
\label{fig:R-x1-x2}
\end{figure}

In Fig. \ref{fig:R-x1-x2}, we show the obtained $R$ and $\rho_0/n_{\rm imp}$
for various on-site potential $I$ as function of $x_1$, 
in the case of (a)(b) $x_2=0$ and (c)(d) $x_2=x_1$.
In the former case, the obtained $R$ is very small for $I=0$,
while it quickly increases for finite $I$.
In contrast, $R$ in the latter case is large even for $I=0$.
In both cases, the residual resistivity is very small for $x_1>0$,
since the magnitude of the hopping integral is increased locally.
Therefore, the $s_\pm$-wave state
is strongly suppressed by the hopping integral inhomogeneity
in both $x_2=0$ and $x_2=x_1$ cases,
except for a special case $I=x_2=0$.

%strongly depends on the detail of model parameters $\Delta t^{(i)}$.
%More systematic analysis would be required
%to achieve a complete understanding of this issue.

%%%%%%%%%%%%%%%%%%%%%%%%%%%%%%%%%
%%%%%%%%%%%%%%%%%%%%%%%%%%%%%%%%%

%%%%%%%%%%%%%%%%%%%%%%%%%
\subsection{impurity-induced non-local orbital order}
\label{sec:Inoue}

To study the effect of a very wide-range impurity potential,
we analyze the effect of the impurity-induced non-local 
orbital order (NL-OO) derived in Ref. \cite{Inoue} on the $s_\pm$-wave state.
%and show that the obtained $\rho_{0}^{\rm cr}$ and $n_{\rm imp}^{\rm cr}$
%are much smaller than their experimental values.
When this impurity-induced NL-OO is formed,
the residual resistivity per 1\% impurity atoms increases to
$\sim50 \ [\mu\Omega{\rm cm}]$,
because of the enlarged impurity scattering cross section \cite{Inoue}.
This fact would resolve the problem that
the residual resistivity derived from Nakamura parameter
is very small expect for Zn-impurity atom.

Recent discovery of ``electronic nematic transition'' 
in the tetragonal phase, free from any lattice deformation, has been 
attracting great attention.
For example, in ``detwinned'' Ba(Fe$_{1-x}$Co$_x$)$_2$As$_2$ 
\cite{detwin}
under very small uniaxial pressure ($\sim5$MPa), 
sizable in-plane anisotropy of resistivity 
emerges at $T^*$, which is about 10K$\sim$100K higher than $T_{\rm S}$.
The nematic order is also observed in BaFe$_2$(As,P)$_2$
by the magnetic torque measurement 
\cite{Matsuda}.
In Ref. \cite{Inoue}, 
the authors discussed the impurity-induced electronic nematic phase
using the mean-field approximation in real space,
by introducing the quadrupole interaction $g_{\rm quad}$.
When orbital fluctuations develop, local impurity potential induces
NL-OO with $C_2$ symmetry,
actually reported by STM/STS autocorrelation analyses \cite{Davis,Song}.
The large cross section of the NL-OO
gives giant residual resistivity,
When $C_2$ nanostructures are aligned along $a$-axis,
the in-plane anisotropy of resistivity reaches $40$\%,
consistently with experiments \cite{detwin}.

%%%%%%%%%%%%%%%%%%%%%%%%%%%%%%%%%
\begin{figure}[!htb]
\includegraphics[width=0.9\linewidth]{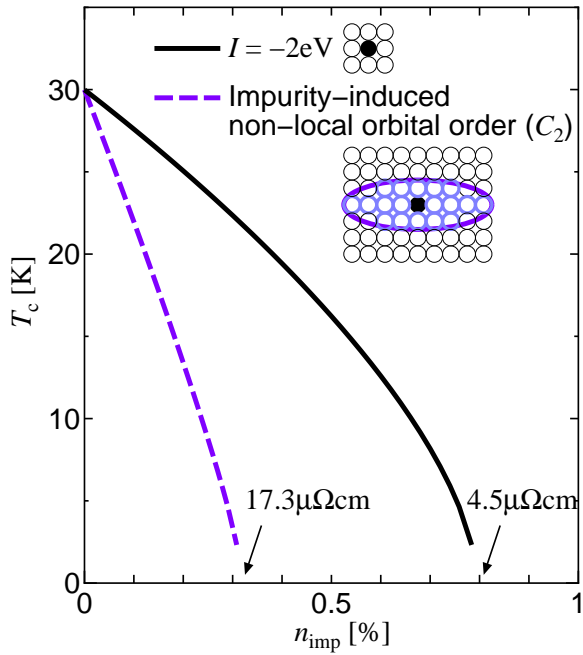}
\caption{
(Color online)
Obtained $T_{\rm c}$ of the $s_\pm$-wave state
in the presence of impurity-induced NL-OO obtained for $I=-2$ 
given in Ref. \cite{Inoue}.
}
\label{fig:Inoue}
\end{figure}
%%%%%%%%%%%%%%%%%%%%%%%%%%%%%%%%%

In Fig. \ref{fig:Inoue},
we show the obtained $T_{\rm c}$ as function of $n_{\rm imp}$
for the (i) on-site impurity potential $I=-2$
and (ii) impurity-induced $C_2$ orbital order obtained by the 
mean-field approximation in real space for $I=-2$ and $g_{\rm quad}=0.218$,
shown in Fig. 1 (c) of Ref. \cite{Inoue}.
In the case (ii), the critical impurity concentration is 
just $n_{\rm imp}^{\rm cr}=0.3z^{-1}\ [\%]$,
which is about one-third of $n_{\rm imp}^{\rm cr}$ for the case (i).
The corresponding in-plane averaged resistivity is
$\rho_0^{\rm cr}=17.3z^{-1} [\mu\Omega{\rm cm}]$,
which is about four times $\rho_0^{\rm cr}$ for the case (i).
That is,  $\rho_0^{\rm cr}$ is enlarged while $n_{\rm imp}^{\rm cr}$ is 
reduced when the impurity-induced NL-OO is realized.
Note that the enlarged $\rho_0^{\rm cr}$ is still one order of
magnitude smaller than experimental values
($300\sim500\ [\mu\Omega{\rm cm}]$).

Therefore, impurity-induced $C_2$ orbital order
should cause strong suppression of the $s_\pm$-wave SC state.
Similar behaviors (larger $\rho_0^{\rm cr}$ and smaller $n_{\rm imp}^{\rm cr}$) 
are also realized by impurity-induced short-range AF order
in nearly AF metals, such as under-doped cuprates \cite{GVI}.
We consider that this impurity-induced NL-OO
will also suppress the $s_{++}$-wave state moderately,
because of the suppression in the density-of-states (pseudo-gap formation) 
%as well as orbital fluctuations are reduced in the presence of
due to the orbital (short-range) order.
Under $T_{\rm c}$, the impurity-induced NL-OO will cause the 
``Swiss cheese hole state'' in the $s_{++}$-wave state.

%%%%%%%%%%%%%%%%%%%%%%%%%
\subsection{derivation of $|T_{\rm inter}|$ and $|T_{\rm intra}|$}

We have introduced various non-local impurity potentials, and
studied the impurity effect on the $s_\pm$ wave state 
driven by the pairing interaction in eq. (\ref{eqn:V}).
We find the relation $R=2.5z \sim 5z \  [{\rm K}/\mu\Omega{\rm cm}]$ 
holds for $3d$- and $4d$-impurity atoms,
whereas $R= 1z \sim 2z \ [{\rm K}/\mu\Omega{\rm cm}]$ in a special model.
%The obtained $R$ is comparable to $R_{I=\infty}=3.6 \ K/{\mu\Omega{\rm cm}}$
%for $3d$-impurity atoms, while $R\lesssim R_{I=\infty}/3$ is realized 
%in the $V_1$-model.
To understand these numerical results,
we analyze the averaged $T$-matrix between FS$\a$ and FS$\b$:
\begin{eqnarray}
|T_{\rm inter}|^2 \equiv \frac{1}{6} \sum_{\a}^{\rm e-FS}\sum_{\b}^{\rm h-FS}
\frac{\int_{\rm FS\a} d\k \int_{\rm FS\b} d\k' |T_{\a\k,\b\k'}|^2}
{\int_{\rm FS\a} d\k \int_{\rm FS\b} d\k' 1},
 \\
|T_{\rm intra}|^2 \equiv \frac{1}{13}
\left( \sum_{\a,\b}^{\rm e-FS}+\sum_{\a,\b}^{\rm h-FS} \right)
\frac{\int_{\rm FS\a} d\k \int_{\rm FS\b} d\k' |T_{\a\k,\b\k'}|^2}
{\int_{\rm FS\a} d\k \int_{\rm FS\b} d\k' 1} .
\end{eqnarray}
Then, the averaged ratio
between inter-pocket and intra-pocket scattering amplitude would be
$t^2\equiv |T_{\rm inter}|^2/|T_{\rm intra}|^2$.

%%%%%%%%%%%%%%%%%%%%%%%%%%%%%%%%%
\begin{figure}[!htb]
\includegraphics[width=0.99\linewidth]{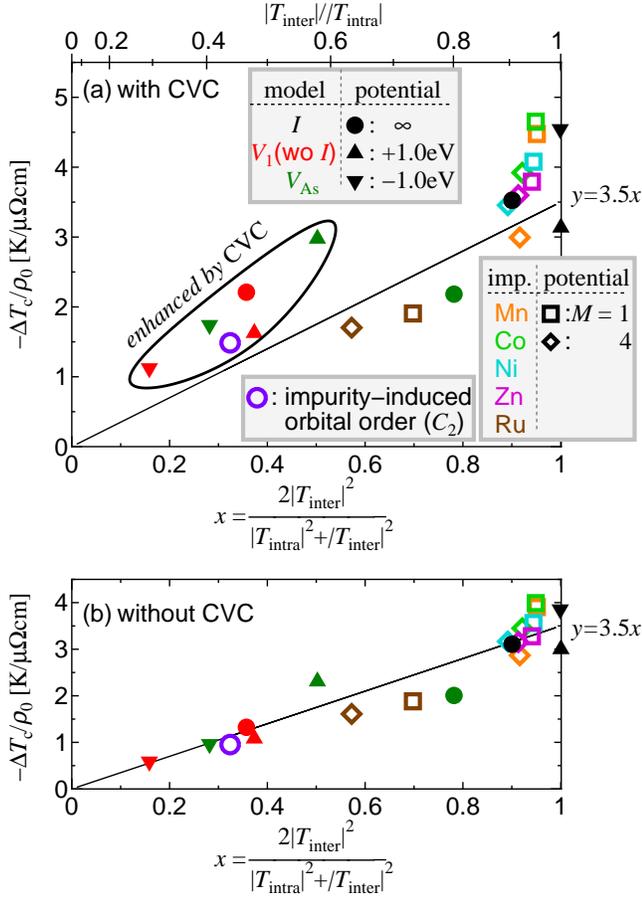}
\caption{
(Color online)
Obtained $R$'s as function of 
$x=2|T_{\rm inter}|^2/(|T_{\rm inter}|^2+|T_{\rm intra}|^2)$
in the case of (a) with CVC and (b) without CVC, for
$3d$- and $4d$-impurity atoms as well as 
$I$-model, $V_{\rm As}$-model, and $V_1$-model without $I$:
Note that the last potential is unrealistic.
$|T_{\rm inter}|=|T_{\rm intra}|$ corresponds to $x=1$.
The proportional relation $R \propto x$ becomes worse by including the CVC,
especially for the $V_1$-model without $I$ and $V_{\rm As}$-model.
%since $\rho_0$ is much smaller than $\rho_0^{\rm no \ CVC}$ 
%for these impurities.
}
\label{fig:R-Tmat}
\end{figure}
%%%%%%%%%%%%%%%%%%%%%%%%%%%%%%%%%

Figure \ref{fig:R-Tmat} (a)
shows the values of $R$ in various impurity models as function of 
$x\equiv 2t^2/(1+t^2)=2|T_{\rm inter}|^2/(|T_{\rm inter}|^2+|T_{\rm intra}|^2)$.
%where $x$ is the ratio of inter-pocket scattering amplitude
%among the total scattering. 
The CVC for the conductivity is taken into account correctly.
It is found that $R$ is approximately scaled by $x$
for various kinds of impurities.
This result is naturally understood since
the $s_\pm$-wave state is suppressed by the inter-FS scattering,
whereas both inter- and intra-FS scattering contribute to
$\rho_0 \propto \gamma$,
as understood by eqs. (\ref{eqn:gamma}) and (\ref{eqn:ASigma}).
Therefore, the following relationships would be realize \cite{Chubukov}:
\begin{eqnarray}
 -\Delta T_{\rm c}&\propto& |T_{\rm inter}|^2, 
\\
\rho_0 &\propto& |T_{\rm inter}|^2+|T_{\rm intra}|^2 .
\end{eqnarray}
We will discuss this issue in more detail in Sec. \ref{sec:T-dis}.

Another important finding in  Fig. \ref{fig:R-Tmat} (a) is that 
the value of $t=|T_{\rm inter}|/|T_{\rm inter}|$ in the five-orbital model 
is approximately independent of the impurity potential strength
(in both Born and unitary regimes),
for all the impurity models studied here.
This fact means that the ``constant ${\hat I}_{\rm b}$-model
can be applicable only for Born impurities,
as we will discuss in Sec. \ref{sec:Senga-dis}.

In Fig. \ref{fig:R-Tmat} (a), the values of $R$ for 
$V_1$- and $V_{\rm As}$-models apparently deviate from the $y=3.5x$ line.
This fact originates from the large contribution from the CVC,
which becomes important when the 
impurity potential has finite scattering cross section, 
since the forward impurity scattering
is correctly subtracted by taking the CVC into account.
In fact, as shown in Fig. \ref{fig:R-Tmat} (b),
the relation $R\propto x$
is apparently improved by neglecting the CVC.
Therefore, we should take the CVC into account 
to obtain quantitatively reliable value of $R$,
especially for wide-range impurity potentials.

%%%%%%%%%%%%%%
\section{Discussions}

In the previous section,
we calculated the impurity effect on the $s_\pm$-wave state
for various non-local impurity models.
When we use the impurity model parameters 
obtained by a recent first principle study by Nakamura {\it et al}.,
the $s_\pm$-wave state is fragile against impurities,
consistently with the previous theoretical study for the 
on-site impurity model by Onari and Kontani \cite{Onari-imp}.
This result is qualitatively unchanged for different non-local 
impurity models, when the parameters are realistic.
Here, we present more detailed discussions.

%%%%%%%%%%%%%%%
\subsection{Estimation of the value of $V_1$}
\label{sec:V1-dis}
% V_1<0.2
In Fig. \ref{fig:R-pot} (a),
we have shown that $R=-\Delta T_{\rm c}/\rho_0$ for the $V_1$-impurity model
is as small as $1z \ [{\rm K}/\mu\Omega{\rm cm}]$,
%is about one-fourth of $R$ for $3d$-impurity atoms shown in
%Fig. \ref{fig:Nakamura-R-M}.
%The value  $R\sim 1z \ [{\rm K}/\mu\Omega{\rm cm}]$
which is still much larger than experimentally observed relations
$R \lesssim 0.1 \ [{\rm K}/\mu\Omega{\rm cm}]$ in single crystals
\cite{Li2,Nakajima}.
Moreover, as shown in Fig. \ref{fig:R-V1} (a),
$R$ is strongly enlarged by introducing the no-site potential $I$,
which should be much larger than $V_1$ in magnitude in real impurities.

By introducing $V_1$
the electron number at the NN site of the impurity center
is changed by $\Delta N_{\rm NN} \sim -2 \times V_1$ for $|V_1|\ll1$,
according to the analysis of the 
present five-orbital tight-binding model.
%That is, the total change of the four NN sites is
%$4\Delta N_{\rm NN} \sim \pm 1$ when $V_0=\pm 0.1$.
However, according to the first principle study \cite{Zawa},
$|\Delta N_{\rm NN}|$ would be at most $0.1\sim0.2$,
meaning that $|V_1|<0.1$ due to the strong screening effect in real compounds.
Therefore, the effect of $V_1$ would be negligible 
in the study of the impurity effect in Fe-based superconductors.
%as recognized in Fig. \ref{fig:R-V1} (a).

%%%%%%%%%%%%%
\subsection{Why $|T_{\rm inter}|\ll |T_{\rm intra}|$ in the $V_1$-model?}
\label{sec:T-dis}

We have shown in Fig. \ref{fig:R-Tmat}
that the relation $|T_{\rm inter}|\ll |T_{\rm intra}|$ 
holds in some non-local impurity potentials.
To understand the reason,
we consider a orbital-less square lattice model for simplicity.
In the Born approximation, the $T$-matrix for the $V_1$-model
is given as 
\begin{eqnarray}
T_{\k,\p}(\e)&=& 4V_1^2 g_{(0,0)}(\e)
\nonumber \\
&&+ 4V_1^2 g_{(1,1)}(\e) (\cos k_x \cos p_y + \cos k_y \cos p_x)
 \nonumber \\
&&+2V_1^2 g_{(2,0)}(\e) (\cos (k_x-p_x) + \cos (k_y-p_y)) ,
  \nonumber \\
\label{eqn:T-kp}
\end{eqnarray}
where $g_{(x,y)}(\e)$ is the Green function in real space at ${\bm r}=(x,y)$.
The cosine terms in eq. (\ref{eqn:T-kp})
originate from the non-locality of the impurity potential.
For the intra hole-pocket scattering ($\k=\p={\bm 0}$),
eq. (\ref{eqn:T-kp}) gives 
$T_{\rm intra}=4V_1^2 (g_{(0,0)}+2g_{(1,1)}+g_{(2,0)})$.
On the other hand, for the scattering between hole- and electron-pockets
($\k=(0,0)$ and $\p=(\pi,0)$), eq. (\ref{eqn:T-kp}) gives 
$T_{\rm inter}=4V_1^2 g_{(0,0)}$.
Therefore, the relation $|T_{\rm inter}|\ll |T_{\rm intra}|$ 
would be possible when $g_{(0,0)}\sim g_{(1,1)}\sim g_{(2,0)}$.

%%%%%%%%%%%%%
\subsection{Comparison with the constant ${\hat I}_{\rm b}$ model}
\label{sec:Senga-dis}

In this paper, we analyzed the impurity effect based on 
the realistic five-orbital model with non-local impurity potentials.
On the other hand, more simple two-band model with 
constant ${\hat I}_{\rm b}$ impurity potential 
has been frequently used 
\cite{Senga1,Senga2}.
Here, we discuss both the usefulness and limitations of the latter model,
in which the impurity potential in the band-basis ($a$, $b$) is given as
$\hat{I}^{\rm b}=
\left(\begin{array}{cc}I_{\rm b} & I'_{\rm b} \\ 
I'_{\rm b} & I_{\rm b} \end{array}\right)$.
If we assume $I_{\rm b}$ and $I'_{\rm b}$  are constant, then we obtain
\begin{eqnarray}
t_{{\rm constant}-I_{\rm b}}^2=\frac{|T_{ab}|^2}{|T_{aa}|^2}
= \frac{u^2}{1+\pi^2 N(0)^2 I_{\rm b}^2(1-u^2)} , 
\label{eqn:r-constantI}
\end{eqnarray}
where 
$T_{ab}$ ($T_{aa}$) is inter-band (intra-band) $T$-matrix.
Here, $u \equiv I_{\rm b}'/I_{\rm b}$, 
and $N(0)$ is the density-of-states for each band.
In the Born regime $\pi N(0)I_{\rm b}\ll1$,
then $t_{{\rm constant}-I_{\rm b}}^2\approx u^2$ holds, and therefore
the $s_{\pm}$-wave state is fragile against impurities except when $u\ll1$.

In the unitary regime $\pi N(0)I_{\rm b} \gg1$,
the $s_{\pm}$-wave state is robust against impurities unless $|u|=1$
since eq. (\ref{eqn:r-constantI}) decreases in proportion to $I_{\rm b}^{-2}$.
%In the supplement of Ref. \cite{Hirsch-imp},
%$\rho_0^{\rm cr}$ increases to $\approx2000\ [\mu\Omega{\rm cm}]$ 
%for $\pi N(0)\sim5$ at $r=0.5$.
However, this result is totally changed in the five-orbital model,
in which ${\hat I}_{\rm b}(\k,\k')=
{\hat U}^\dagger(\k)\cdot {\hat W}(\k,\k')\cdot{\hat U}(\k')$ 
is momentum-dependent:
%once the realistic momentum-dependence of ${\hat I}_{\rm b}$ 
%is taken into account.
Once ${\hat I}_{\rm b}$ is $\k$-dependent,
then eq. (\ref{eqn:r-constantI}) does not hold 
as proved in Ref. \cite{Onari-imp}.
Instead, the relation 
\begin{eqnarray}
t_{I_{\rm b}(\k,\k')}^2\sim u^2
\end{eqnarray}
holds for 
{\it all the impurity models studied here even in the unitary regime}, 
as recognized by the numerical analysis in Fig. \ref{fig:R-Tmat}.
Therefore, the constant ${\hat I}_{\rm b}$ model 
is applicable to Fe-based superconductors only for Born impurities.

%%%%%%%%%%%%%%%%%%%%%%%%%%%%%%%%%
\begin{figure}[!htb]
\includegraphics[width=0.85\linewidth]{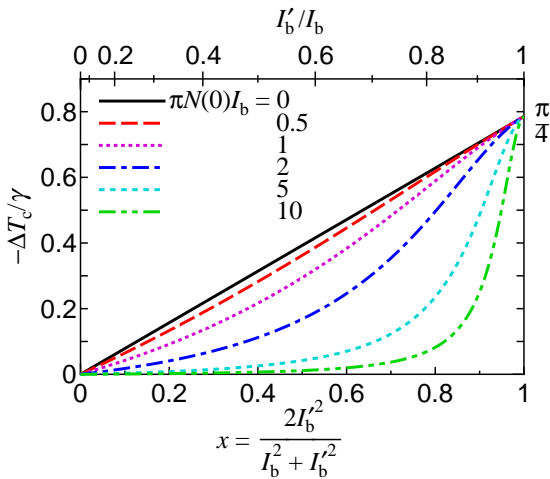}
\caption{
(Color online)
Obtained $-\Delta T_{\rm c}/\gamma$ of the $s_\pm$-wave state
in the two-band constant ${\hat I}_{\rm b}$ model studied in Ref. \cite{Senga1}.
%We put the DOS at the Fermi energy is $N(0)=1$ for both bands.
The horizontal axis is
$x\equiv 2u^2/(1+u^2)=2I_{\rm b}'^2/(I_{\rm b}'^2+I_{\rm b}^2)$.
$I_{\rm b}'=I_{\rm b}$ corresponds to $x=1$.
}
\label{fig:Senga}
\end{figure}
%%%%%%%%%%%%%%%%%%%%%%%%%%%%%%%%%

Figure \ref{fig:Senga} shows the 
$-\Delta T_{\rm c}/\gamma$ in the two-band constant ${\hat I}_{\rm b}$ model 
as function of $x\equiv 2u^2/(1+u^2)=2I_{\rm b}'^2/(I_{\rm b}'^2+I_{\rm b}^2)$,
in case that the $s_\pm$-wave state due to inter-band repulsion is realized.
In this model, $-\Delta T_{\rm c}/\gamma$  is given by eq. (12)
of Ref. \cite{Senga1}.
Note that $\rho_0\propto \gamma$.
We see that the relationship $-\Delta T_{\rm c}/\gamma \propto x$
holds only in the Born regime $\pi N(0)I_{\rm b} \ll1$.
However, the relation $-\Delta T_{\rm c}/\gamma \ll x$ is realized
in the unitary and intermediate regimes except for $|u|\sim1$.

%Based on the constant ${\hat I}_{\rm b}$-model, 
%Ref. \cite{Hirsch-imp} mentioned that 
%the $s_{\pm}$-wave state with $T_{\rm c0}=30$K disappears for 
%$\rho_0^{\rm cr}\approx 100 \ (1000)\ [\mu\Omega{\rm cm}]$ at $u=0.5 \ (0.2)$
%in the intermediate regime $\pi N(0)I_{\rm b} \sim2.5$.
%In the five-orbital, however, 
%$\rho_0^{\rm cr}$ with CVC for $t=u$ is much smaller.
%For example, in case of the impurity-induced NL-OO in Fig. \ref{fig:Inoue},
%in which $t\approx 0.4$ is realized,
%we obtain $\rho_0^{\rm cr}=17.3 z^{-1}\sim50 \ [\mu\Omega{\rm cm}]$.

Based on the constant ${\hat I}_{\rm b}$-model, 
Ref. \cite{Hirsch-imp} mentioned that 
the $s_{\pm}$-wave state with $T_{\rm c0}=30$K disappears for 
$\rho_0^{\rm cr}\approx 100 \ (1000)\ [\mu\Omega{\rm cm}]$ at $u=0.5 \ (0.2)$
in the intermediate regime $\pi N(0)I_{\rm b} \sim2.5$,
and tried to explain the experimental small impurity effect on $T_{\rm c}$
based on the $s_\pm$-wave scenario assuming that $u\ll0.5$.
However, we obtain $t=u\sim1$ 
for the realistic impurity models of $3d$-impurity atoms
as shown in Fig. \ref{fig:R-Tmat}.
In addition, 
$\rho_0^{\rm cr}$ given in Ref. \cite{Hirsch-imp} is very overestimated 
for $u\ll 0.5$, since they apply the constant ${\hat I}_{\rm b}$-model
to the intermediate regime and the CVC for the resistivity is dropped.
In fact, we obtain $\rho_0^{\rm cr}=17.3 z^{-1}\sim50 \ [\mu\Omega{\rm cm}]$
for the impurity-induced NL-OO in Fig. \ref{fig:Inoue};
$t=u\sim0.4$ in this case would be the lower limit for 
realistic impurity models for Fe-based superconductors.

%we obtain $\rho_0^{\rm cr}=17.3 z^{-1}\sim50 \ [\mu\Omega{\rm cm}]$.

In this paper, we studied the effect of in-plane impurities.
In the case of ``out-of-plane'' impurities, the radius of the 
impurity potential $\phi$ could be much longer than the lattice spacing
$a_{\rm Fe-Fe}$ \cite{Graser-long-range-imp-pot}.
In this case, the relation $u=|I_{b}'/I_{b}|\ll1$ might be realized 
because of the suppression of large angle scattering,
%will be small when $|k_{\rm in}-k_{\rm out}| > \pi/\phi$
%Since the impurity potential in the momentum space $V(k,k')$ 
%will be small when $|k-k'| > \pi/\phi$, the ratio $u=|I_{b}'/I_{b}|$ could be 
%suppressed when $\phi\gg a_{\rm Fe-Fe}$, 
and then the $s_{\pm}$-wave state would be robust against impurities.
Therefore, study of the out-of-plane impurities
would not be useful to distinguish between $s_{++}$- and $s_{\pm}$-wave states.

%Thus, the impurity-induced nodal $s$ $\rightarrow$ $s_\pm$ crossover discussed in 
%Ref. \cite{Hirsch-imp}, which can be realized if $|T_{ab}|^2/|T_{aa}|^2\ll1$,
%would not be easy to be realized experimentally.

%However, at present we have no information
%on the rondomness introduced by irradiations.

%%%%%%%%%%%%%%%%%%%%%%%%%%%%%%%%%%%%%
\section{summary}

Based on the realistic five-orbital model
for Fe-based superconductors,
we have presented a systematic study of the impurity effect on 
the $s_\pm$-wave SC state.
We studied impurity model parameters for $3d$- and $4d$-impurity atoms
obtained by a recent first principle study \cite{Nakamura},
in addition to various non-local impurity models.
The obtained values of $R=-\Delta T_{\rm c}/\rho_0$ for $n_{\rm imp}\ll1$
as function of 
$x=2|T_{\rm inter}|^2/(|T_{\rm inter}|^2+|T_{\rm intra}|^2)$
are summarized in Fig. \ref{fig:R-Tmat}.
According to the Abrikosov-Gorkov theory,
$R$ is essentially independent of $T_{\rm c0}$.

The main results are shortly summarized as follows:
\\ \\
(i) For $3d$-impurity atoms, $|T_{\rm inter}|/|T_{\rm intra}|\gtrsim 0.9$
is realized, and we obtain $R=3z \sim 5z \ [{\rm K}/\mu\Omega{\rm cm}]$,
which is comparable to $R_{I=\infty}=3.6z \ [{\rm K}/\mu\Omega{\rm cm}]$
for the on-site model with $I=\infty$.
Also, $\rho_0^{\rm cr} \sim 5z^{-1} \ [\mu\Omega{\rm cm}]$ 
for $T_{\rm c0}=30$K.
\\ \\
(ii)  For Ru-impurity atoms, $|T_{\rm inter}|/|T_{\rm intra}|=0.6\sim0.7$
%(or $x=0.26\sim0.33$) 
is realized, and both $R$ and $1/\rho_0^{\rm cr}$ are
about half of those for $3d$-impurity atoms.
\\ \\
(iii) For impurity-induced NL-OO model,
which gives a very wide-range impurity potential,
we obtain $|T_{\rm inter}|/|T_{\rm intra}|\sim 0.4$ and
$R\sim 1.5z \ [{\rm K}/\mu\Omega{\rm cm}]$.
Similar result is obtained for the $V_{\rm As}$-model
and $V_1$-model with finite $I$.
%For $V_1$-model (without $I$) and $V_{\rm As}$-model,
%$|T_{\rm intra}|/|T_{\rm inter}|= 0.3\sim 0.6$ 
%(or $x=0.08\sim 0.25$) 
%is realized, and we obtain $R=1z \sim 3z \ [{\rm K}/\mu\Omega{\rm cm}]$.
\\ \\
(iv) The CVC is important to obtain reliable $R$ and $\rho_0^{\rm cr}$:
$R$ is approximately doubled by CVC for $V_1$- and 
$V_{\rm As}$-impurity models, for example.
\\ \\
(v) In case we use the ``constant ${\hat I}_{\rm b}$ impurity model''
by putting $|I_{\rm b}'/I_{\rm b}|=|T_{\rm inter}|/|T_{\rm intra}|$,
we have to keep in the Born regime $\pi N(0) I_{\rm b}\ll1$,
especially when $|I_{\rm b}'/I_{\rm b}|\ll1$.
\\

The abovementioned results are essentially
consistent with the previous theoretical study for the 
on-site impurity model by Onari and Kontani \cite{Onari-imp}.
%This result is essentially unchanged for different 
%non-local impurity models with realistic parameters:
%$|I|\gg |V_1|$ and $|V_1|<0.1$.
Experimentally, the critical value of $\rho_0$
for the disappearance of $T_{\rm c0}\approx 30$K is
$300\sim500\ [\mu\Omega{\rm cm}]$,
which means that the averaged mean-free-path 
is comparable to the lattice spacing ($\sim0.3$nm)
as discussed in Ref. \cite{Onari-imp}.
Therefore, experimentally observed robustness of $T_{\rm c}$
against various kinds of impurities
in optimally-doped Fe-based superconductors
\cite{Sato-imp,Nakajima,Li2,Paglione}
would be consistent with the $s_{++}$-wave state.

In this paper, we have introduced only the repulsive pairing 
interaction due to spin fluctuations.
When both spin and orbital fluctuations strongly develop,
inter-orbital attractive interaction and intra-orbital repulsive 
interaction coexist.
This situation is naturally expected near the orthorhombic phase, 
and is actually reproduced by the SC-VC theory 
based on the Hubbard model \cite{Onari-SCVC}.
Then, the present study indicates that impurity-induced crossover 
form $s_\pm$-wave to $s_{++}$-wave states would be realized 
in some Fe-based superconductors
\cite{Kontani-RPA,Hirsch-cross}.

%\vspace{5mm}

\acknowledgments
We are grateful to M. Sato, Y. Kobayashi, J. Li,
Y. Matsuda, T. Shibauchi, and M. Nakajma for valuable discussions. 
This study has been supported by Grants-in-Aid for Scientific 
Research from MEXT of Japan.
Numerical calculations were partially performed using 
the Yukawa Institute Computer Facility.

%%%%%%%%%%%%%%%%%%%%%%%%%
%references
%%%%%%%%%%%%%%%%%%%%%%%%


\begin{thebibliography}{99}


\bibitem{Hosono}
Y. Kamihara, T. Watanabe, M. Hirano, and H. Hosono:
% Y. Kamihara {\it et al}.: 
J. Am. Chem. Soc. {\bf 130}, 3296 (2008).

\bibitem{Hashimoto}
K. Hashimoto, et al.: Phys. Rev. Lett. {\bf 102} (2009) 017002.

\bibitem{Tanatar}
M. A. Tanatar, et al.: Phys. Rev. Lett. {\bf 104} (2010) 067002.

\bibitem{Hirsch-rev}
P. J. Hirschfeld, M. M. Korshunov, and I. I. Mazin, 
Rep. Prog. Phys. {\bf 74} (2011) 124508.

\bibitem{ARPES-Shen}
M. Yi, D. H. Lu, J.-H. Chu, J. G. Analytis, A. P. Sorini, A. F. Kemper, S.-K. Mo,
R. G. Moore, M. Hashimoto, W. S. Lee, Z. Hussain, T. P. Devereaux, I. R. Fisher, Z.-X. Shen,
PNAS {\bf 108}, 6878 (2011).
%arXiv:1011.0050.

\bibitem{Fernandes}
R.M. Fernandes, L. H. VanBebber, S. Bhattacharya, P. Chandra, 
V. Keppens, D. Mandrus, M.A. McGuire, B.C. Sales, A.S. Sefat, 
and J. Schmalian, Phys. Rev. Lett. {\bf 105}, 157003 (2010) 

\bibitem{Yoshizawa}
%M. Yoshizawa, R. Kamiya, R. Onodera, Y. Nakanishi, K. Kihou, H. Eisaki, and C. H. Lee, arXiv:1008.1479.
M. Yoshizawa {\it et al.}, 
Phys. Soc. Jpn. {\bf 81}, 024604 (2012).

\bibitem{Goto}
%T. Goto, R. Kurihara, K. Araki, K. Mitsumoto, M. Akatsu, Y. Nemoto, 
%S. Tatematsu, and M. Sato, 
T. Goto {\it et al.},
J. Phys. Soc. Jpn. {\bf 80}, 073702 (2011).

\bibitem{Kuroki}
K. Kuroki, S. Onari, R. Arita, H. Usui, Y. Tanaka, H. Kontani, and H. Aoki,
%K. Kuroki {\it et al}., 
Phys. Rev. Lett. {\bf 101}, 087004 (2008).
%K. Kuroki {\it et al}., Phys. Rev. B {\bf 79}, 224511 (2009).

\bibitem{Mazin}
I. I. Mazin, D.J. Singh, M.D. Johannes, and M.H. Du:
Phys. Rev. Lett. {\bf 101} (2008) 057003.

\bibitem{Hirschfeld}
S. Graser, G. R. Boyd, C. Cao, H.-P. Cheng, P. J. Hirschfeld, 
and D. J. Scalapino, Phys. Rev. B {\bf 77}, 180514(R) (2008).

\bibitem{Chubukov}
A. V. Chubukov, D. V. Efremov, and I. Eremin,
Phys. Rev. B {\bf 78}, 134512 (2008). 

\bibitem{Kontani-RPA}
H. Kontani and S. Onari, Phys. Rev. Lett. {\bf 104}, 157001 (2010).

\bibitem{Saito-RPA}
T. Saito, S. Onari, and H. Kontani,
Phys. Rev. B {\bf 82}, 144510 (2010).
 
\bibitem{Saito-RPA2}
T. Saito, S. Onari, and H. Kontani,
Phys. Rev. B {\bf 83}, 140512(R) (2011) 

\bibitem{Onari-SCVC}
S. Onari and H. Kontani, Phys. Rev. Lett. {\bf 109}, 137001 (2012) 

\bibitem{Ohno-SCVC}
Y. Ohno, M. Tsuchiizu, S. Onari and H. Kontani,
J. Phys. Soc. Jpn. {\bf 82} (2013) 013707 

\bibitem{Hanaguri}
T. Hanaguri, S. Niitaka, K. Kuroki, and H. Takagi,
Science {\bf 328}, 474 (2010)

\bibitem{christianson}
A. D. Christianson, E. A. Goremychkin, R. Osborn, S. Rosenkranz, M. D. Lumsden, C. D. Malliakas,
I. S. Todorov, H. Claus, D. Y. Chung, M. G. Kanatzidis, R. I. Bewley, and T. Guidi, Nature {\bf 456}, 930 (2008).

\bibitem{keimer}
D. S. Inosov, J. T. Park, P. Bourges, D. L. Sun, Y. Sidis,
	A. Schneidewind, K. Hradil, D. Haug, C. T. Lin, B. Keimer, and V. Hinkov
,Nature Physics {\bf 6}, 178 (2010)

%\bibitem{zhao}
%J. Zhao, L. -P. Regnault, C. Zhang, M. Wang, Z. Li, F. Zhou, Z. Zhao, P. Dai, Phys. Rev. B {\bf 81}, 180505 (2010).

\bibitem{tate}
S. Tatematsu, Y. Yasui, T. Moyoshi, K. Motoya, K. Kakurai, and M. Sato,
 to be published in J. Phys. Soc. Jpn.

\bibitem{Sato-imp}
A. Kawabata, S. C. Lee, T. Moyoshi, Y. Kobayashi, and M. Sato,
J. Phys. Soc. Jpn. {\bf 77} (2008) Suppl. C 103704;
M. Sato, Y. Kobayashi, S. C. Lee, H. Takahashi, E. Satomi, and Y. Miura,
J. Phys. Soc. Jpn. {\bf 79} (2009) 014710;
S. C. Lee, E. Satomi, Y. Kobayashi, and M. Sato,
J. Phys. Soc. Jpn. {\bf 79} (2010) 023702.

\bibitem{Nakajima}
Y. Nakajima, T. Taen, Y. Tsuchiya, T. Tamegai, H. Kitamura, and T. Murakami,
Phys. Rev. B {\bf 82}, 220504 (2010).

%\bibitem{Li1}
%J. Li, Y. Guo, S. Zhang, S. Yu, Y. Tsujimoto, H.Kontani, K. Yamaura, 
%and E. Takayama-Muromachi:
%Phys. Rev. B {\bf 84}, 020513(R) (2011). 

\bibitem{Li2}
J. Li, Y. F. Guo, S. B. Zhang, J. Yuan, Y. Tsujimoto, X. Wang, 
C. I. Sathish, Y. Sun, S. Yu, W. Yi, K. Yamaura, 
E. Takayama-Muromachiu, Y. Shirako, M. Akaogi, and H. Kontani,
Phys. Rev. B {\bf 85}, 214509 (2012)

\bibitem{Paglione}
K. Kirshenbaum, S. R. Saha, S. Ziemak, T. Drye, and  J. Paglione,
%K. Kirshenbaum {\it et al}.,
Phys. Rev. B {\bf 86}, 140505 (2012)

\bibitem{Onari-resonance}
%S. Onari {\it et al.},
S. Onari, H. Kontani, and M. Sato,
Phys. Rev. B {\bf 81}, 060504(R) (2010);
S. Onari and H. Kontani, Phys. Rev. B {\bf 84}, 144518 (2011)

\bibitem{Senga1}
Y. Senga and H. Kontani: J. Phys. Soc. Jpn. {\bf 77} (2008) 113710;

\bibitem{Bang}
Y. Bang, H.Y. Choi and H. Won, Phys. Rev. B {\bf 79}, 054529 (2009).

\bibitem{Senga2}
Y. Senga and H. Kontani, New J. Phys. {\bf 11}, 035005 (2009). 

\bibitem{Onari-imp}
S. Onari and H. Kontani, 
Phys. Rev. Lett. {\bf 103} (2009) 177001.

\bibitem{FCZhang}
Z.J. Yao, W.Q. Chen, Y.K. Li, G.H. Cao, H.M. Jiang, Q.E. Wang,
Z.A. Xu and F.C. Zhang, Phys. Rev. B {\bf 86}, 184515 (2012) 

\bibitem{Hirsch-imp}
 Y. Wang, A. Kreisel, P. J. Hirschfeld and V. Mishra,  arXiv:1210.7474

\bibitem{Nakamura}
K. Nakamura, R. Arita and H. Ikeda,
Phys. Rev. B {\bf 83}, 144512 (2011) 

\bibitem{Kawamata}
T. Kawamata, E. Satomi, Y. Kobayashi, M. Itoh, and  M. Sato,
J. Phys. Soc. Jpn. {\bf 80} (2011) 084720

\bibitem{Canfield1}
P. C. Canfield, S. L. Budko, Ni Ni, J. Q. Yan, and A. Kracher, 
Phys. Rev. B {\bf 80}, 060501(R) (2009).

\bibitem{Canfield2}
N. Ni, A. Thaler, J.Q. Yan, A. Kracher, E. Colombier,
S. L. Budko, P. C. Canfield, and S. T. Hannahs, 
Phys. Rev. B {\bf 82}, 024519 (2010).

\bibitem{Ideta}
 S. Ideta, T. Yoshida, I. Nishi, A. Fujimori, Y. Kotani, K. Ono, 
Y. Nakashima, S. Yamaichi, T. Sasagawa, M. Nakajima, K. Kihou, 
Y. Tomioka, C. H. Lee, A. Iyo, H. Eisaki, T. Ito, S. Uchida and R. Arita,
Phys. Rev. Lett. {\bf 110}, 107007 (2013)

\bibitem{ROP}
H. Kontani, Rep. Prog. Phys. {\bf 71}, 026501 (2008) 

\bibitem{Inoue}
 Y. Inoue, Y. Yamakawa and  H. Kontani,
Phys. Rev. B {\bf 85}, 224506 (2012) 

\bibitem{Ikeda}
H. Ikeda, K. Nakamura and R. Arita, unpublished.

\bibitem{Zawa}
 H. Wadati, I. Elfimov and G. A. Sawatzky,
Phys. Rev. Lett {\bf 105}, 157004 (2010) 

\bibitem{Kemper}
A. F. Kemper, C. Cao, P. J. Hirschfeld, and H.-P. Cheng, Phys. Rev. B {\bf 80}, 104511 (2009);
{\it ibid}, Phys. Rev. B {\bf 81}, 229902(E).

\bibitem{detwin}
J.-H. Chu, J. G. Analytis, K. D. Greve, P. L. McMahon, Z. Islam, 
Y. Yamamoto, and I. R. Fisher,
Science {\bf 329}, 824 (2010);
J. J. Ying, X. F. Wang, T. Wu, Z. J. Xiang, R. H. Liu, Y. J. Yan,
A. F. Wang, M. Zhang, G. J. Ye, P. Cheng, J. P. Hu, and X. H. Chen,
 Phys. Rev. Lett. {\bf 107}, 067001 (2011) 

\bibitem{Matsuda}
 S. Kasahara, H.J. Shi, K. Hashimoto, S. Tonegawa, Y. Mizukami, 
T. Shibauchi, K. Sugimoto, T. Fukuda, T. Terashima, A. H. Nevidomskyy
and Y. Matsuda,
Nature {\bf 486}, 382 (2012) 

\bibitem{Davis}
T.-M. Chuang, M.P. Allan, J. Lee, Y. Xie, N. Ni, S.L. Budko, G.S. Boebinger, 
P.C. Canfield and J.C. Davis,
Science {\bf 327}, 181 (2010)

\bibitem{Song}
C.-L. Song {\it et al.}, 
Science {\bf 332}, 1410 (2011). 

\bibitem{GVI}
H. Kontani and M. Ohno, Phys. Rev. B {\bf 74}, 014406 (2006)

\bibitem{Graser-long-range-imp-pot}
S. Graser,P. J. Hirschfeld, L.Y. Zhu and T. Dahm, 
Phys. Rev. B {\bf 76}, 054516 (2007).

\bibitem{Hirsch-cross}
D. V. Efremov, M. M. Korshunov, O. V. Dolgov, A. A. Golubov
and  P. J. Hirschfeld,
Phys. Rev. B {\bf 84}, 180512(R) (2011) 
 
\end{thebibliography}
\end{document}